# Free-form micro-optics enabling ultra-broadband low-loss fiber-to-chip coupling

**Shaoliang Yu[1†], Luigi Ranno[1†], Qingyang Du[1], Samuel Serna[1,2*], Colin McDonough[3], Nicholas Fahrenkopf[3], Tian Gu[1,4*], and Juejun Hu[1,4]**

[1]*Department of Materials Science & Engineering, Massachusetts Institute of Technology, Cambridge, MA, USA*
[2]*Department of Physics, Photonics and Optical Engineering, Bridgewater State University, Bridgewater, MA, USA*
[3]*The Research Foundation for State University of New York, State University of New York Polytechnic Institute, Albany, NY, USA*
[4]*Materials Research Laboratory, Massachusetts Institute of Technology, Cambridge, MA, USA*

† These authors contributed equally to this work.
*ssernaotalvaro@bridgew.edu, gutian@mit.edu

## ABSTRACT

Efficient fiber-to-chip coupling has been a major hurdle to cost-effective packaging and scalable interconnections of photonic integrated circuits. Conventional photonic packaging methods relying on edge or grating coupling are constrained by high insertion losses, limited bandwidth density, narrow band operation, and sensitivity to misalignment. Here we present a new fiber-to-chip coupling scheme based on free-form reflective micro-optics. A design approach which simplifies the high-dimensional free-form optimization problem to as few as two full-wave simulations is implemented to empower computationally efficient design of high-performance free-form reflectors while capitalizing on the expanded geometric degrees of freedom. We demonstrated fiber array coupling to waveguides taped out through a standard foundry shuttle run and backend integrated with 3-D printed micro-optics. A low coupling loss down to 0.5 dB was experimentally measured at 1550 nm wavelength with a record 1-dB bandwidth of 300 nm spanning O to U bands. The coupling scheme further affords large alignment tolerance, high bandwidth density and solder reflow compatibility, qualifying it as a promising optical packaging solution for applications such as wavelength division multiplexing communications, broadband spectroscopic sensing, and nonlinear optical signal processing.

## INTRODUCTION

Photonic integrated circuits (PICs), the optical analog of electronic integrated circuits, are emerging as a mainstream technology in place of traditional bulky optical systems. Volume manufacturing leveraging standard semiconductor foundry processes have been driving steady cost down scaling and performance improvement of PICs, underpinning their widespread deployment in applications including communications, sensing, medical imaging, computing, quantum information, and navigation. Photonic packaging, an essential step to optically interface PICs with other components (in particular optical fibers) to form a functional system, is however increasingly becoming the bottleneck for PIC applications, since it often requires customized tools with low throughput and accounts for up to 80% of photonic module costs[1,2].

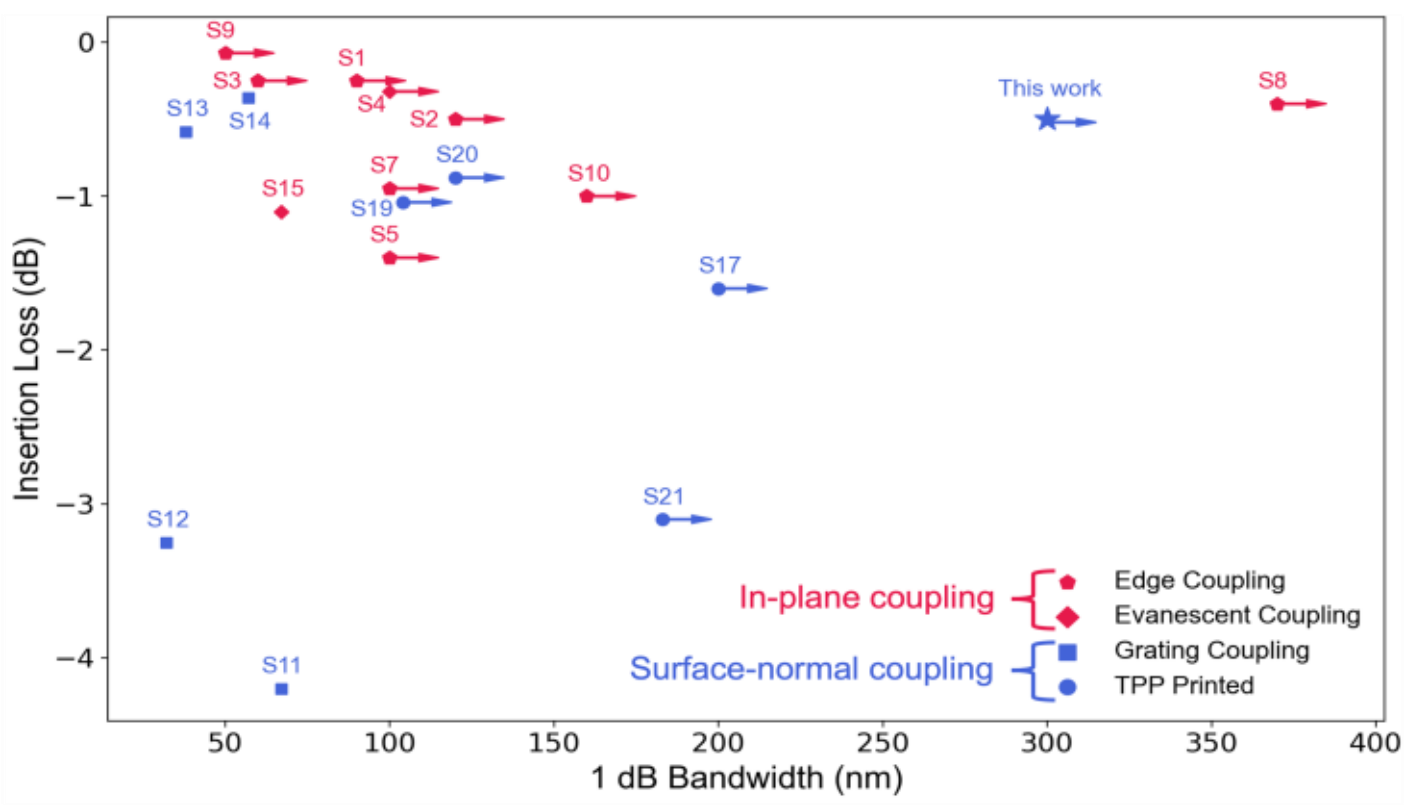

**Fig. 1. Summary of insertion loss and 1-dB bandwidth performances of fiber-to-chip coupling techniques demonstrated to date: the arrows indicate that the 1-dB bandwidth of the coupler exceeds the measurement wavelength range. Corresponding literature references and detailed quantitative information are tabulated in Supplementary Section 1.**

The "packaging bottleneck" results from the more stringent alignment tolerance, wavelength dependence, and sensitivity to optical losses inherent to optical coupling between different components, making it difficult to harness the existing electronic packaging infrastructure to significantly lower cost and boost throughput. To put this in context, Table S1 compares different fiber-to-chip coupling approaches, including edge and grating coupling which are commonplace in traditional photonic packaging as well as emerging coupling techniques[3,4]. Edge coupling requires access to chip facets/edges, which not only precludes wafer-level testing but also limits bandwidth density due to the large 1-D fiber pitch. On the other hand, while diffractive grating coupling is compatible with wafer-level characterization, it is narrow-band and unsuitable for wavelength division multiplexing (WDM) or broadband sensing. Figure 1 summarizes the insertion loss and 1-dB bandwidth performances of various fiber-to-chip coupling schemes demonstrated to date. Notably, while sub-1dB coupling losses have been reported for both schemes, silicon photonic foundries typically specify coupling losses of 1.5 dB (edge) and 3 – 4 dB (grating)[5,6], since special structures such as embedded metal mirrors or multilayer poly-Si overlay required for high-efficiency designs are not available in standard foundry runs.

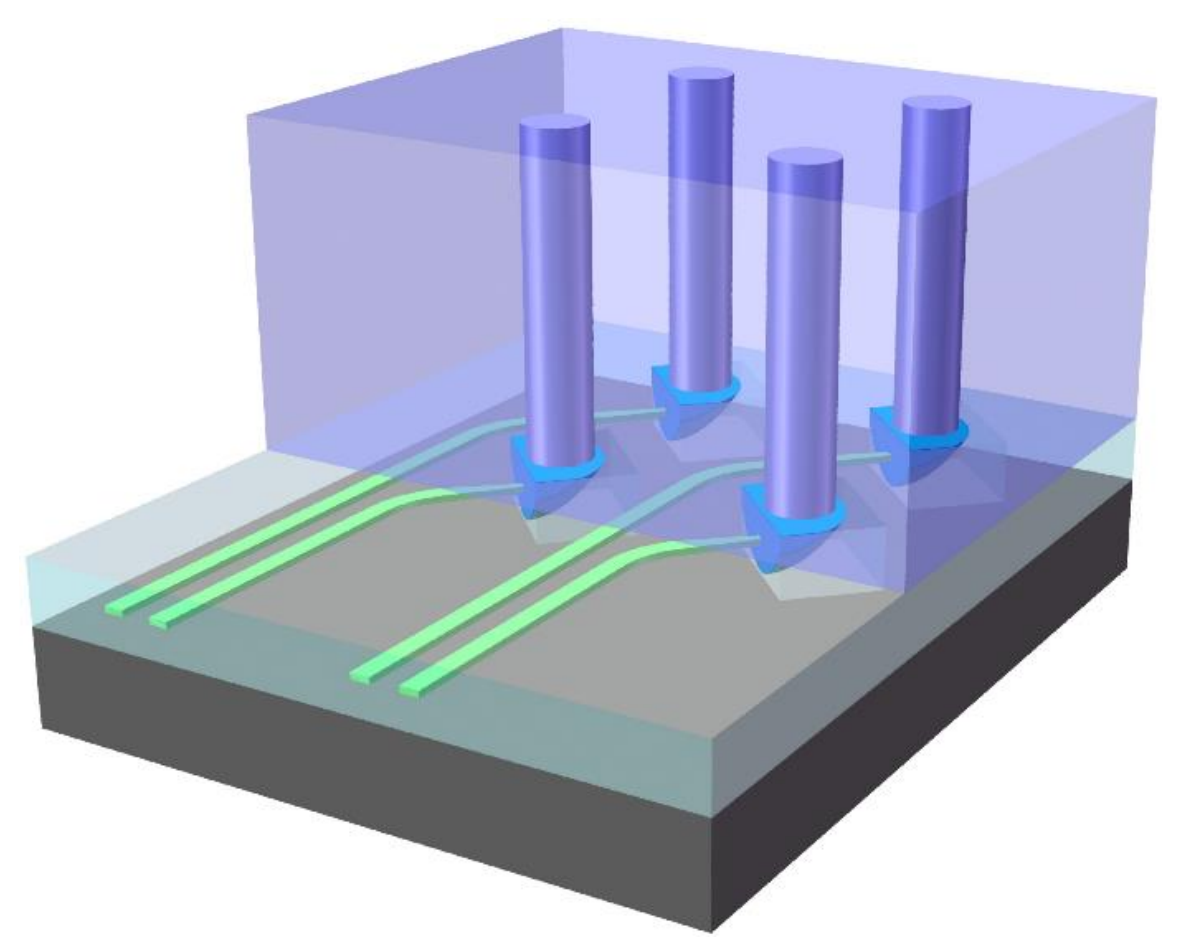

**Fig. 2. 3-D rendering of the free-form micro-optical fiber-to-chip coupling structure**

In this paper, we report a universal photonic coupling scheme based on free-form micro-optical reflectors as illustrated in Fig. 2. Unlike diffraction or refraction, optical reflection is inherently wavelength agnostic and incurs vanishingly low optical loss in the total internal reflection (TIR) regime. The reflectors serve to redirect and shape the waveguide output, enabling high-density surface-normal coupling to 2-D fiber arrays with low losses and high tolerance to misalignment. Fabrication of the micro-optical reflectors makes use of two photon polymerization (TPP), an additive 3-D printing technique offering sub-wavelength resolution ideal for free-form micro-optics definition[7–14]. Original work from several groups has successfully implemented TPP for optical coupling between passive waveguides[15], lasers[16], free space[17], and fibers[18–24]. Besides acting as an optical coupling element, TPP-written structures have also been used as optomechanical supports to enhance alignment accuracy during photonic assembly[25,26]. These pioneering results validate TPP as a viable technique for robust and versatile photonic packaging that is maturing into the commercial domain[27]. Nonetheless, these early demonstrations still leave ample room for improvements on key aspects such as design, fabrication and integration. The minimal fiber-to-chip coupling loss remains at ~ 1 dB and above. Moreover, the aforementioned fiber-to-chip coupling designs operate exclusively with unclad waveguides and therefore entail custom top cladding stripping steps, which are usually not part of standard photonic foundry processes. Last but not least, geometric fidelity of the TPP-written coupler structures has not been vetted upon heat treatment, which is necessary to establishing solder reflow compatibility of TPP-based packaging techniques.

The research reported herein is poised to circumvent these limitations. The couplers are designed for waveguides embedded inside dielectric claddings, the norm for foundry-processed PICs. We experimentally demonstrated a coupling loss of 0.5 dB, representing the lowest loss figure reported for surface-normal couplers at 1550 nm wavelength. The high efficiency benefits from the free-form reflector design, which allows full utility of the geometric degrees of freedom to precisely shape the wavefront and thereby facilitating fiber-to-waveguide mode matching. Our coupler further achieves ultra-wide band operation with a record 1-dB bandwidth of 300 nm and coupling losses consistently below 2 dB across the entire second and third telecom windows (O, E, S, C, L, and U bands). Finally, we proved that the couplers' geometric fidelity and hence coupling efficiency are not compromised upon heat treatment up to 250 ˚C, the upper bound of typical non-lead solder reflow temperature. The combination of low loss, broadband operation, high bandwidth density, as well as wafer-scale testing and solder reflow compatibility qualify our approach as a promising optical interfacing solution for applications such as WDM communications, low coherence interferometry, quantum optics, on-chip optical tweezing, and spectroscopic sensing.

## WAVEFRONT-BASED FREE-FORM MICRO-OPTICAL COUPLER DESIGN

Free-form optics, which are characterized by surfaces with no axis of rotational invariance[28], confer on-demand wavefront control taking advantage of the large number of accessible degrees of freedom. The geometric flexibility, however, poses a design challenge if the full design space is to be explored. Design of free-form optics largely relies upon parameterization of the surface geometry (e.g. polynomial expansion) and subsequent multidimensional optimization. The computation overhead for solving such optimization problems in the 3-D full-vectorial regime is often rather cumbersome.

Here we propose a design approach based on wavefront interference which simplifies the free-form design problem into minimally two finite-difference time-domain (FDTD) simulations. The approach is illustrated in Fig. 3, where our task is to search for a reflective surface geometry that transforms the in-plane waveguide mode to the out-of-plane fiber mode in an orthogonal direction, or performs other mode matching functions. The Fermat's principle specifies that the total optical path length connecting the two modes' wavefronts while passing through the reflector must be stationary with respect to variations of the path, i.e.

$$\varphi_1 + \varphi_2 + \varphi_3 = \varphi_{tot}, \tag{1}$$

where $\varphi_1$ and $\varphi_2$ correspond to the phase delays associated with light traveling from the waveguide to the reflector and from the reflector to the fiber, respectively, $\varphi_3$ is the phase delay incurred by TIR at the reflector, and $\varphi_{tot}$ is a constant. As we show in the Supplementary Section 2, $\varphi_3$ is only weakly dependent on the light incidence location on the reflector and thus can be treated as a constant and will henceforth be absorbed into $\varphi_{tot}$.

Next, we consider the configuration in Fig. 3b which models light output from the fiber in the absence of the waveguide and the reflector. In the figure, the red curve (marked as W) labels the wavefront of light exiting from the fiber after propagating for some distance in free space. We denote the phase delay from fiber to W as $\varphi_4$, which is a constant independent of the light path. Now if we back propagate the light from W to the fiber, the phase delay from W to a point on the reflector is given by $\varphi_4$ - $\varphi_2$. Equation (1) translates to:

$$\varphi_1 - (\varphi_4 - \varphi_2) = \varphi_{tot} - \varphi_4 \equiv 2N\pi, \tag{2}$$

Here $\varphi_{tot}$ - $\varphi_4$ is a constant and we set it to integral multiples of $2\pi$. This is possible since we can always tune the position of W such that $N$ becomes an integer. Equation (2) implies that when output from the waveguide and the back propagated fiber mode from W co-propagate in space, the loci of constructive interference (given by Eq. 2) coincide with a group of reflector surface shapes that satisfy the Fermat's principle. The optimal reflector geometry is then uniquely determined by choosing a surface from the group which maximizes overlap integral between the waveguide output and the back propagated fiber mode over the surface.

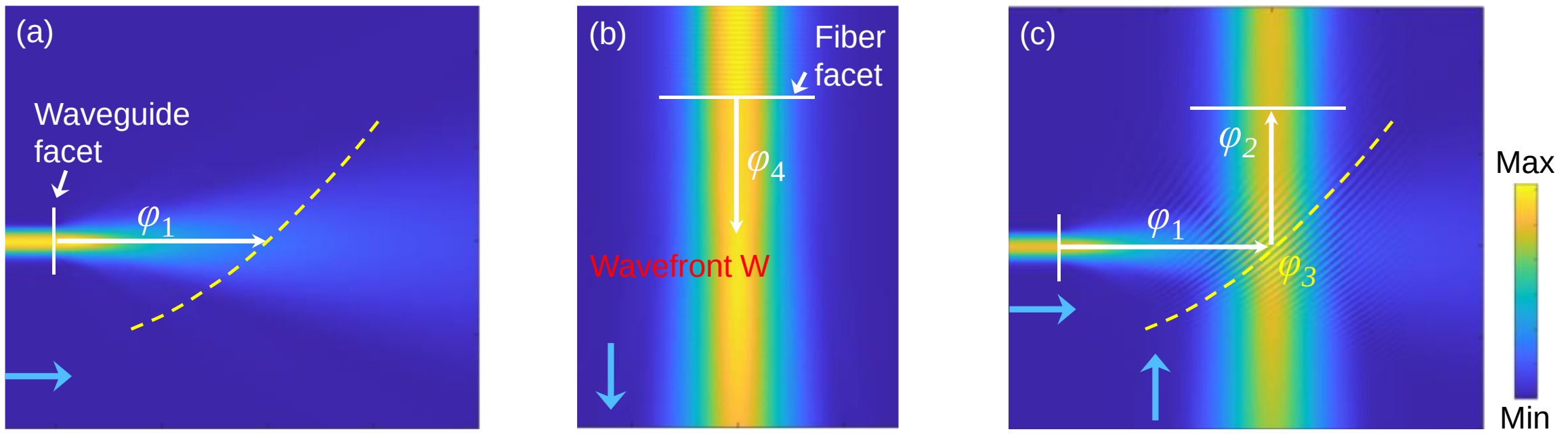

**Fig. 3. The free-form reflector design method based on wavefront interference : (a, b) FDTD-simulated field intensity profiles with separate optical outputs from (a) the waveguide and (b) the fiber, where the arrows denote phase delays; (c) concurrent modeling of the waveguide output and the back-propagated fiber output. Fringes due to interference between the two beams correspond to traces having a constant optical path linking the waveguide and the fiber (labeled with the dotted curve).**

The design approach outlined here is deterministic and requires only two 3-D FDTD simulations: the fiber mode propagation without the waveguide, and co-propagation of the waveguide output and the back propagated fiber mode. Once the second simulation is complete, we extract the loci of maximum local intensity points using a spiral trace search algorithm (Supplementary Section 3). The electromagnetic field components on the surface coordinates are subsequently used to evaluate the overlap integral. In addition to providing a computationally efficient route for optimization of waveguide coupling to standard single-mode fibers, we further show that the method is also applicable to designing unconventional coupler structures. As an example highlighting versatility of the approach, we designed a micro-optical coupler between a SiN waveguides and a polarization-maintaining endlessly single-mode photonic crystal fiber, with the latter having an irregular mode profile without circular symmetry (Supplementary Section 4). The design claims a peak coupling efficiency of 91.8% (0.37 dB insertion loss) and a 1-dB bandwidth exceeding 500 nm. Such a design potentially enables efficient coupling of broadband light (e.g. supercontinuum) between on-chip waveguides and fibers while maintaining single-mode operation with octave-spanning bandwidth.

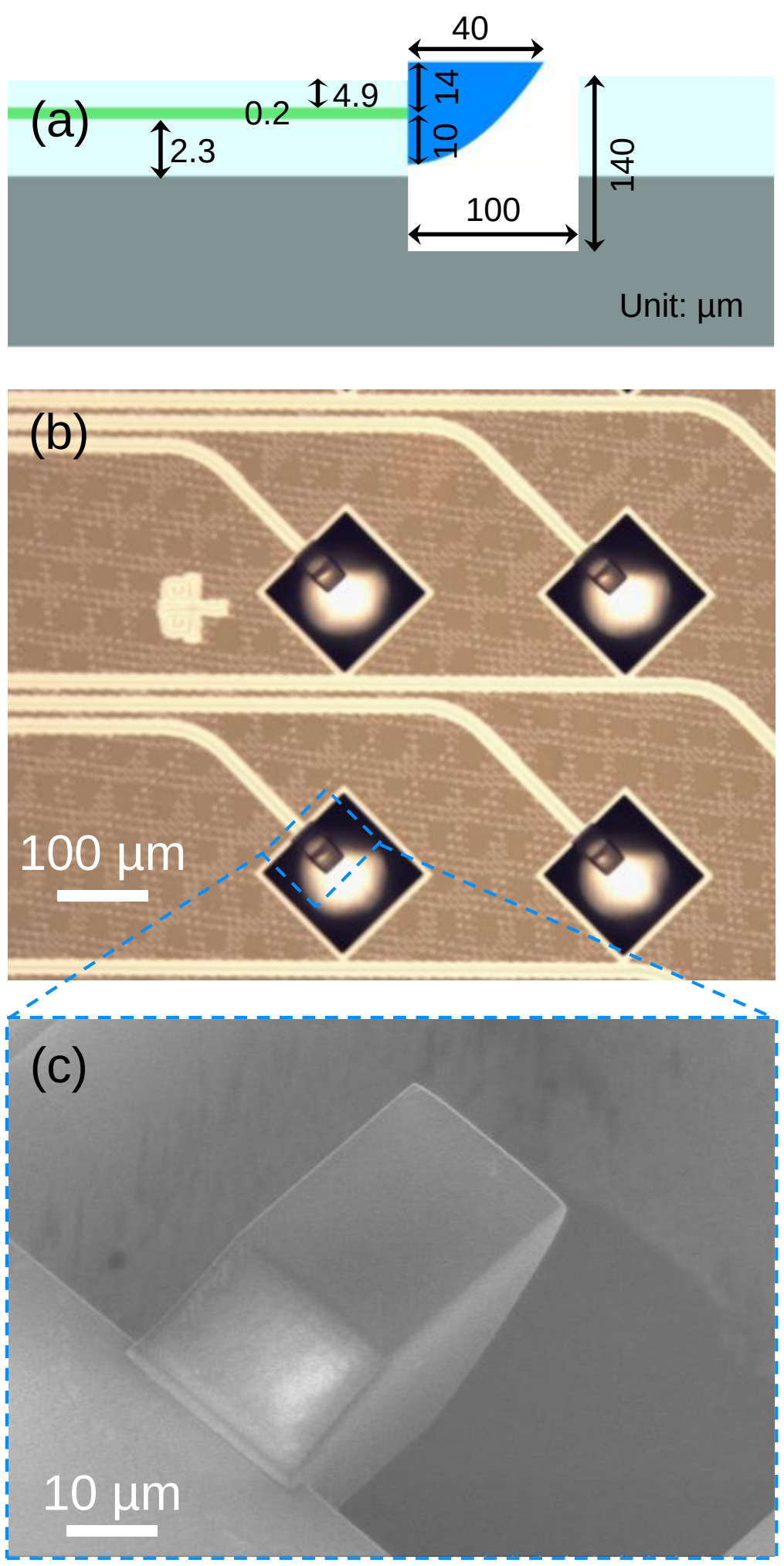

**Fig. 4. (a) Schematic cross-section of the micro-reflector (not drawn to scale). The dimensions (in μm) are marked. (b) Top-view optical micrograph of the foundry-processed chip. (c) Close-up view of the coupler structure.**

## DEVICE FABRICATION AND CHARACTERIZATION

As a proof of concept, here we demonstrate integration of the free-form couplers with foundry-processed SiN waveguides, while the reflective coupler configuration is equally applicable to silicon-on-insulator waveguides with similar low-loss, ultra-broadband performance (Supplementary Section 5). The SiN waveguides were fabricated on an American Institute for Manufacturing Integrated Photonics (AIM Photonics) multi-project wafer (MPW) shuttle run[29] Figure 4a schematically depicts the cross-sectional structure of the fabricated device. The SiN waveguides, designed to support a single TE-polarization mode, are 1,500 nm in width and are adiabatically tapered to a width of 360 nm at the end tips. The recesses, which have a depth of 140 μm and act to expose the waveguide's vertical facets for coupler attachment, are defined during the dicing trench patterning step via reactive ion etching as part of the standard shuttle run process flow. The micro-optical couplers were subsequently fabricated using TPP, and details of the fabrication process are elaborated in Methods. Figure 4b presents a top-view optical micrograph of the finished chip and Fig. 4c furnishes a close-up view of the coupler structure. The couplers exhibit excellent adhesion to the waveguide facets with no sign of mechanical failure. The coupler surfaces feature a low root-mean-square roughness of 18 nm, implying negligible optical losses due to roughness scattering.

Figure 5 summarizes FDTD modeling results on the fabricated coupler structure, which is optimized to couple TE-polarized light from a SiN waveguide to an SMF-28 fiber. Leveraging the capability of free-form reflective surfaces in precise wavefront shaping, the intensity profile of the beam exiting from the coupler closely matches that of the fiber mode, which accounts for the exceptionally low coupling loss: < 1 dB throughout the second and third telecom windows (1260 to 1675 nm). The in-plane alignment tolerance corresponding to 1-dB loss penalty is ± 2.2 μm, consistent with the fiber mode field diameter. In the out-of-plane direction, the reflected beam is almost collimated, giving rise to negligible (< 0.1 dB) loss penalty over a distance range of 28 μm. The tolerance can be further enhanced by expanding the output beam diameter and writing an additional lensing structure on the fiber facet.

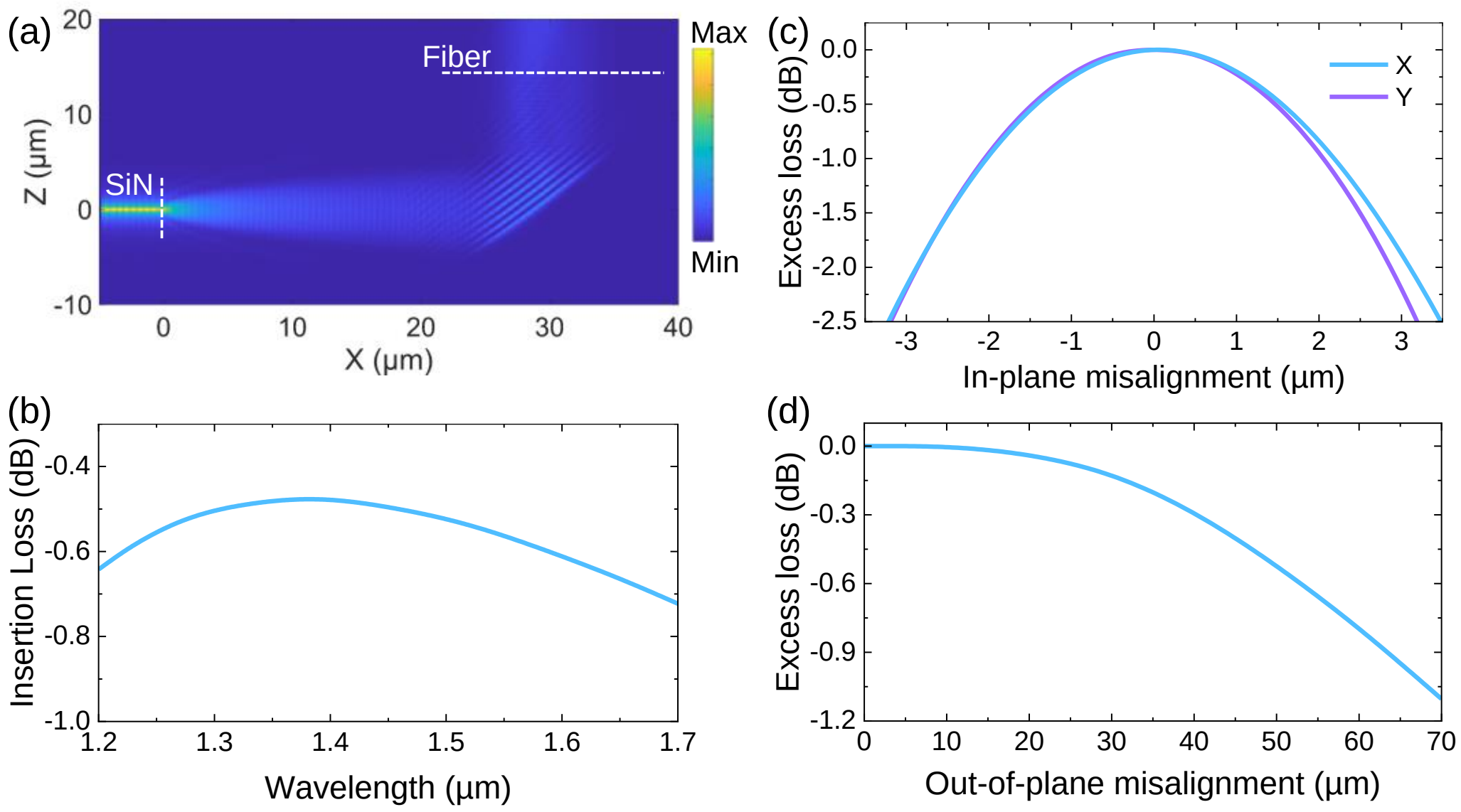

**Fig. 5. The micro-optical coupler modeled using 3-D FDTD: (a) simulated electric field intensity distribution across the waveguide center plane; (b) modeled insertion loss spectrum; and (c, d) alignment tolerances along (c) in-plane and (d) out-of-plane directions.**

Optical waveguides with micro-optical couplers attached to both of their input and output facets were tested using SMF-28 fiber arrays with an inter-fiber pitch of 250 μm, matching that of the couplers on-chip (Fig. 6a). Wavelength-dependent coupling losses were inferred from the total insertion loss of the coupler-waveguide-coupler loop, after subtracting propagation loss contributions from the waveguide. Detailed experimental characterization protocols are discussed in the Methods section. Figure 6b plots the measured loss spectra of the coupler. At 1550 nm wavelength, the coupler boasts a low coupling loss of 0.5 dB consistent with our modeling result. The 1-dB bandwidth covers over 300 nm from 1340 to 1640 nm, whose upper bound is only limited by the accessible wavelength range of our lasers. Over the measured wavelength range from 1260 to 1640 nm, the coupling loss is below 2 dB, enabling the coupler to be used for WDM applications spanning the entire long-wave telecom bands. In fact, we have experimentally observed coupling of visible light from a fiber into and out of the chip through the couplers as well (Fig. 6c), although the coupling efficiency was not quantified given the multi-mode nature of the SiN waveguides at wavelengths below 1260 nm. The slight deviation from the simulation result is attributed to shape distortion of the reflector likely caused by polymer volume shrinkage during crosslinking, as well as overtone absorption of the polymer which accounts for the loss peak at 1380 nm wavelength (Supplementary Section 6). We also verified that the coupling loss remained unchanged after heat treatment at 250 ˚C (Fig. 6d), a key proof validating solder reflow compatibility of the micro-optical coupler technology. The in-plane 1-dB alignment tolerance is ± 2.2 μm (Fig. 6e), commensurate with the positioning accuracy of high-precision passive alignment instruments. In accordance with the simulation result in Fig. 5, the coupler is insensitive to longitudinal offset between the fiber and the coupler with an out-of-plane 1-dB alignment tolerance of 20 μm.

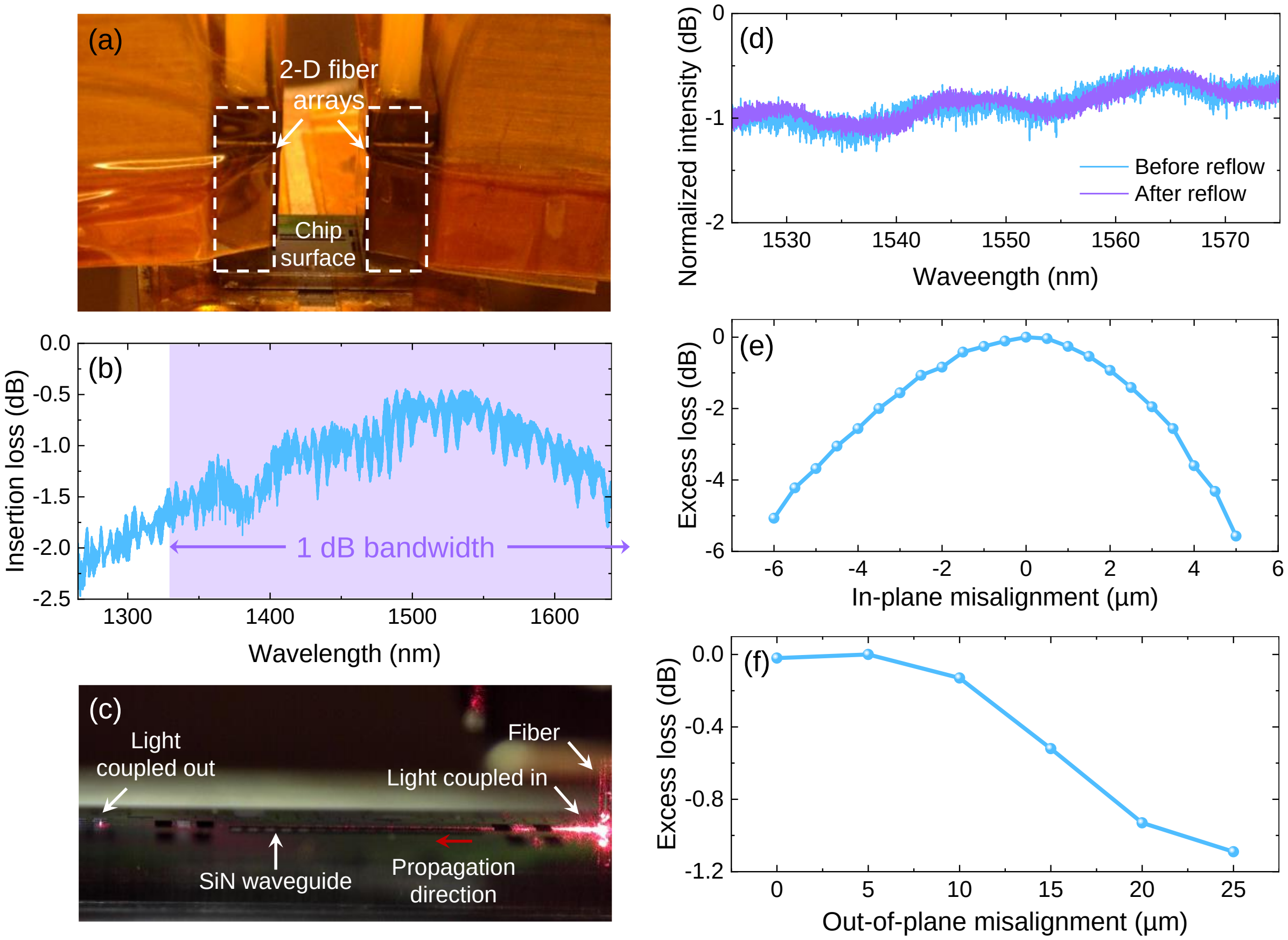

**Fig. 6. (a) A photo of 2-D fiber arrays coupling to on-chip waveguides via the reflective couplers. (b) Measured broadband insertion loss of the coupler. (c) A photo showing visible light coupling into and out of a waveguide via the same coupler designed for 1550 nm wavelength. (d) Transmission spectra before (blue) and after (purple) annealing at 250 ˚C to simulate solder reflow heat treatment. (e, f) Measured (e) in-plane and (f) out-of-plane alignment tolerances.**

## DISCUSSION

The low-loss and broadband micro-optical coupler technology provides a promising optical coupling platform for integrated photonics packaging in applications ranging from WDM communications to broadband spectroscopic sensing. As an example showcasing the potential application of the technology in nonlinear optics, we contrasted the nonlinear characteristics of SiN waveguides terminated with the free-form micro-optical couplers and standard grating couplers (but are otherwise identical). In Fig. 7a, transmission spectra of the waveguides were obtained with femtosecond pulse trains with a repetition rate of 100 MHz, an average power of 80 mW, and a pulse duration of 100 fs (corresponding to a peak power of 8 kW). When the pulses were launched through grating couplers, the resulting spectrum (purple curve) is truncated due to limited bandwidth of the grating couplers. In contrast, the micro-optical coupler preserved spectral information of the pulses (blue curve). Figure 7b compares spectral broadening due to self-phase modulation in SiN waveguides upon injection of 13 nm wide top-hat pulses centered around 1565 nm wavelength, showing that the nonlinear broadening is far more pronounced in waveguides with the micro-optical couplers. This enhanced nonlinearity benefits from the high coupling efficiency and hence higher on-chip optical power, as well as the micro-optical coupler's wide bandwidth which minimizes spectral distortion. Importantly, no sign of optical damage was observed in the micro-optical couplers after pumping using ultrafast pulses with peak powers up to 8 kW. The unique combination of low loss, large bandwidth and high power stability qualifies the micro-optical couplers as a robust optical interface of choice for nonlinear photonic device packaging and characterization.

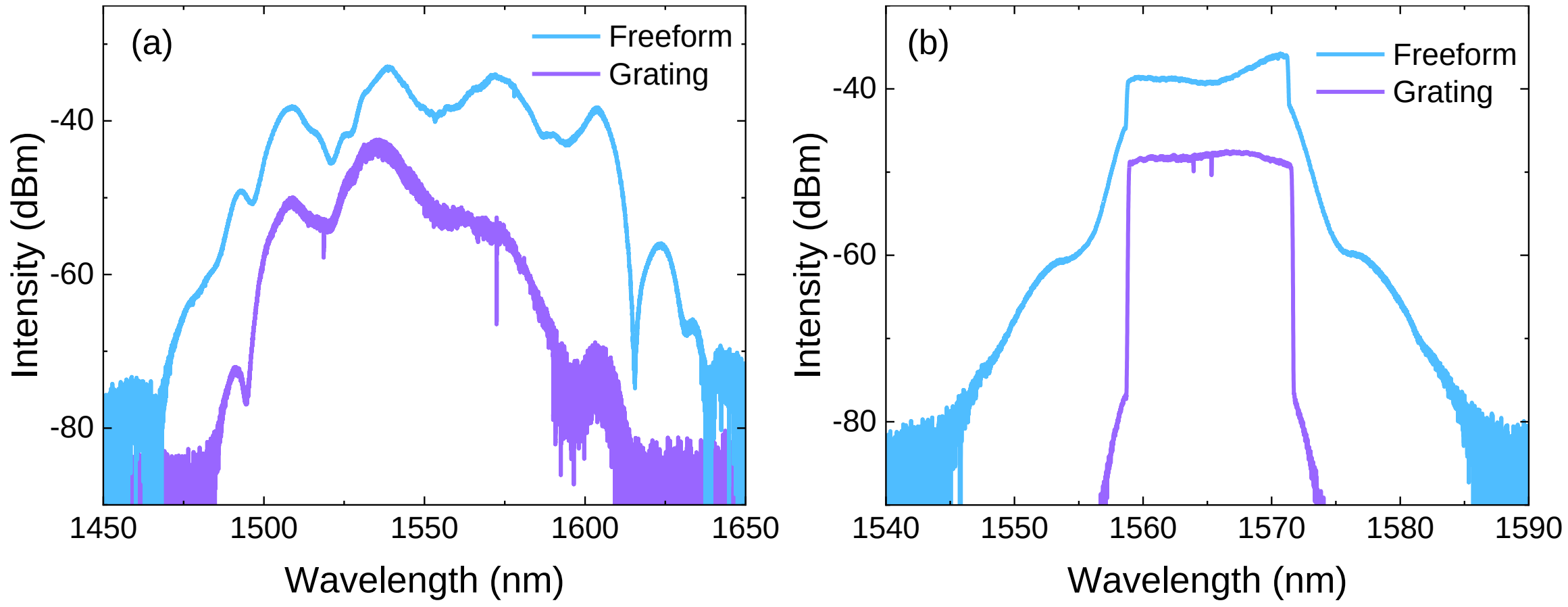

**Fig. 7. Transmission spectra of (a) 100 fs femtosecond pulses with 8 kW peak power; and (b) 13 nm wide top-hat pulses after spectral broadening due to self-phase modulation in SiN waveguides. The pulses were coupled into the waveguides via the free-form micro-optical couplers (blue curves) and grating couplers (purple curves) with identical input power.**

Besides functioning as a fiber-to-chip coupling interface, the micro-optical free-form reflectors can also be applied to enable efficient chip-to-chip and chip-to-interposer coupling. Solder-reflow compatibility of the couplers implies that they can be implemented alongside solder bonds to form flip-chip optical and electrical connections between two chips concurrently with potential self-alignment capability, owing to the large misalignment tolerance of the couplers. An instance of the simultaneous optical and electrical connection design is illustrated in Supplementary Section 7. The free-form reflector further offers a versatile method to control the wavefront and beam shape of light coupling from on-chip waveguides to free space, a feature instrumental to applications such as free space communications, remote sensing, optical manipulation[17], and quantum optics[30].

In summary, we demonstrated a fiber-to-chip coupling scheme leveraging micro-optical reflectors which are integrated on standard foundry-processed photonic chips. The free-form couplers are designed using a wavefront-based, computationally efficient approach which avoids time-consuming parameter search used in traditional free-form optics design. Experimentally fabricated couplers measure a low insertion loss of 0.5 dB at 1550 nm wavelength, a 1-dB bandwidth of > 300 nm, and < 2 dB coupling loss across the O to U telecom bands. Moreover, we have validated that the couplers are solder reflow compatible and can withstand high power pulses with peak power exceeding 8 kW. The same reflective coupler architecture is further applicable to efficient broadband chip-to-chip, chip-to-interposer, and chip-to-free-space coupling. We therefore anticipate that the universal micro-optical coupling interface will find widespread use spanning data communications, spectroscopic sensing, quantum optics, and nonlinear optical signal processing, among many other applications.

## METHODS

**Device fabrication.** The photonic chips were fabricated in a multi-project AIM shuttle run. The trenches used to expose the waveguide facets were defined during the dicing-trench etching step. Subsequently, micro-reflectors made of IP-n162 were printed on the side wall of the pre-defined trenches via TPP using a commercially-available machine (Photonic Professional GT, Nanoscribe GmbH) with the following printing parameters: hatching distance 100 nm, slicing distance of 200 nm, writing speed of 5 mm/s, and average writing power 60 mW (780 nm, 80 MHz, ~100 fs).

**Optical characterization.** The optical performance of the couplers was characterized using a tunable laser with built-in vector analyzer (OVA 5000). The broadband measurements were performed using 3 different tunable lasers from Santec's TSL series, controlled through an optical switch module (OSU-100). The nonlinear measurements were performed using the ELMO High Power-FS laser from Menlo Systems, which was coupled into a standard SMF-28 fiber through a free-space fiber coupler. Coupling was performed using SMF-28 fibers and fiber arrays cleaved at 0°. The chips were tested on an automated probe station (SD-100 from Maple Leaf Photonics).

**Device modeling.** Optical simulations were performed using Ansys' Lumerical Finite Difference Time Domain (FDTD) module. Optimization of the reflector surface was done using a home-written Python script. The extracted point cloud corresponding to the loci of constructive interference was output as 3-D STL file using Solidworks® for FDTD simulation and TPP writing.

### Acknowledgments

This work was supported by ARPA-E under the ENLITENED Program (Award Number: DE-AR0000847). The authors thank Dr. Xiaoming Qiu and Anthony Zhou for their help with the modeling code development.

### Author contributions

S.Y., T.G. and J.H. conceived the device design. S.Y. and L.R. designed, fabricated and characterized the devices. Q.D. assisted with device fabrication. S.S. and L.R. carried out the wide band and nonlinear measurements. C.M. and N.F. performed the foundry fabrication and helped with device design. S.Y., L.R. and J.H. drafted the manuscript. T.G. and J.H. supervised and coordinated the research. All authors contributed to revising the manuscript and technical discussions.

### Competing financial interests

The authors declare no competing financial interests.